\documentclass[pra,aps,twocolumn]{revtex4}
\usepackage{graphicx}
\usepackage{amsbsy}
\usepackage{amsmath,amsthm}
\usepackage{amsfonts}

\newcommand{\nn}{\nonumber \\}

\newcommand{\ket}[1]{|{#1}\rangle}
\newcommand{\bra}[1]{\langle{#1}|}
\newcommand{\Alpha}{{\rm A}}
\newcommand{\E}{{\cal E}}
\newcommand{\I}{\mathbf{i}}

\begin{document}

\title{Implementation of multipartite unitary operations with limited resources}

\author{Dominic W.\ Berry}
\affiliation{Department of Physics, The University of Queensland, Brisbane,
Queensland 4072, Australia}

\begin{abstract}
A general method for implementing weakly entangling multipartite unitary
operations using a small amount of entanglement and classical communication is
presented. For the simple Hamiltonian $\sigma_z\otimes\sigma_z$ this method
requires less entanglement than previously known methods. In addition,
compression of multiple operations is applied to reduce the average
communication required.
\end{abstract}

\maketitle

\section{Introduction}
Much of quantum information processing relies on the ability to perform
operations between subsystems. In the case where these subsystems do not
directly interact, it is necessary to achieve the operation by indirect means.
For example, one could use a third subsystem which interacts with both, or use
entanglement. In the case where the operation is only weakly entangling, it
would be useful to be able to implement it using only a small amount of
entanglement.

A SWAP operation may be applied using two teleportations \cite{tele}, requiring
two ebits of entanglement and two bits of communication in each direction.
Similarly, a CNOT may be applied using one ebit and one bit of communication in
each direction \cite{eisert}. For more general two-qubit unitaries, weakly
entangling unitaries may be implemented using a small amount of entanglement
\cite{cirac}. Ref.\ \cite{cirac} gives a scheme to implement operations of the
form $e^{-i\alpha\sigma_x\otimes\sigma_x}$ using average entanglement of
$5.9793\alpha$. This may also be applied to more general two-qubit unitaries.
Alternative schemes for operations of this form, but allowing a significant
probability of failure are presented in Refs.\ \cite{gross,chen}. A scheme for
Hamiltonians which are a tensor product of self-inverse operators was given in
Ref.\ \cite{zeng}.

Here we present a general method of implementing weakly entangling multipartite
unitary operations using a small amount of entanglement or communication. The
situation considered is that there are $N$ parties, each with a number of
subsystems, and operations performed by the parties are assumed to have
negligible cost. The unitary we wish to implement acts upon one subsystem in
possession of each of the parties, and we quantify the entanglement and
classical communication between the parties.

This paper proceeds as follows. The general problem of implementing operations
using entanglement and communication is presented in Sec.\ \ref{sec:gen}. Then
an efficient method for implementing evolution under multipartite Hamiltonians
of the form $\sigma_z^{\otimes N}$ is presented in Sec.\ \ref{sec:entcon}. A
method of using compression over multiple operations to reduce the communication
required is given in Sec.\ \ref{sec:clascom}. A method of generalising these
results to tensor product Hamiltonians is given in Sec.\ \ref{sec:genten}, and
further generalisations are given in Sec.\ \ref{sec:fur}. Conclusions are given
in Sec.\ \ref{sec:con}.

\section{General methods}
\label{sec:gen} The most general method of implementing an operation using
entanglement and classical communication may be described as follows. The $N$
parties share an entangled resource state, and perform many rounds of the
following process. Each party performs a local unitary followed by a partial
measurement. The parties then communicate classical information about the
measurement results, and repeat the process with local unitaries and
measurements possibly depending on the transmitted information.

In general we consider many implementations of $U$, and regard the scheme as
successful if the minimum fidelity of the output state approaches one as the
number of copies of $U$ approaches infinity. We may define a sufficient rate
vector $\vec R$ as a vector of measures of multipartite entanglement and
classical communication such that, for all $\epsilon>0$ there exists an $n$
such that $n$ implementations of $U$ may be performed using average entanglement
and classical communication no larger than the components of $\vec R$, and
minimum fidelity at least $1-\epsilon$.

The scheme of Cirac, D\"ur, Kraus and Lewenstein (CDKL) \cite{cirac} is of this
form. Joint measurements between an entangled resource state and the target
system are made. The correct measurement result means that the operation is
correctly performed. If failure occurs, another resource state is used, and
measurement success corrects for the previous failure and correctly implements
the unitary. For repeated failure this process is repeated until the operation
which is required may be performed locally.

In the following we improve on the CDKL scheme in three ways: \\
1. We reduce the entanglement consumption. \\
2. We reduce the communication in most directions to be the same as the
entanglement consumption. \\
3. We apply the scheme to more general Hamiltonians.

\section{Entanglement consumption}
\label{sec:entcon} First we give an improved scheme which reduces the
entanglement consumed. We consider the Hamiltonian $\sigma_z^{\otimes N}$ for
$N$ parties, so the unitary we wish to apply is
$U(\alpha)=\exp(i\alpha\sigma_z^{\otimes N})$. This is the multipartite
generalisation of the Hamiltonian considered by CDKL. (We use $\sigma_z$ here
rather than $\sigma_x$, but these are equivalent under local unitaries.)

The $N$ parties use the resource state
\begin{equation}
\ket{\psi(\beta)}=\cos(\beta)\ket{0}^{\otimes N}+i\sin(\beta)
\ket{1}^{\otimes N},
\end{equation}
then apply a four step process. In the following we use the ``stator''
formalism, introduced in Ref.\ \cite{stator}. A stator is a hybrid
state-operator object, and acts upon a state by applying its operator component
to the state, and appending its state component. That is, the stator
\mbox{$\ket{\phi}\otimes U$} acts on $\ket{\psi}$ as
\begin{equation}
(\ket{\phi}\otimes U) \ket{\psi} = \ket{\phi}\otimes (U \ket{\psi}).
\end{equation}
Linearity is used to obtain the action of stators which are sums of terms of the
form $\ket{\phi}\otimes U$. The four steps are as follows: \\
Step 1. Each party applies a controlled-$Z$ operation on the target system. This
yields the stator
\begin{equation}
\cos(\beta)\ket{0}^{\otimes N}\otimes\openone^{\otimes N}
+i\sin(\beta)\ket{1}^{\otimes N}\otimes\sigma_z^{\otimes N}
\end{equation}
Step 2. Parties 1 to $N-1$ apply Hadamard operations followed by computational
basis measurements on their component of the resource state. The resulting
stator is
\begin{equation}
\label{eq:pmsta}
\cos(\beta)\ket{0}\otimes\openone^{\otimes N}
\pm i\sin(\beta)\ket{1}\otimes\sigma_z^{\otimes N},
\end{equation}
where the sign is $+$ if the number of measurement results equal to 1 is even,
and $-$ otherwise. \\
Step 3. Party $N$ performs a correction on their component of the resource state
based on the measurement results. If the number of measurement results equal to
1 is odd, then they perform a $\sigma_z$ operation on the remaining portion of
the resource state. We then obtain the stator with the plus
in Eq.\ \eqref{eq:pmsta}. \\
Step 4. Party $N$ applies a projection measurement on their component of the
resource state. Projection onto the state $\ket\psi=\cos(\gamma)\ket{0}+
\sin(\gamma)\ket{1}$ yields the unitary operation proportional to
\begin{equation}
\label{eq:correct}
\cos(\beta)\cos(\gamma)\otimes\openone^{\otimes N}
+ i\sin(\beta)\sin(\gamma)\otimes\sigma_z^{\otimes N}.
\end{equation}
For failure, projection onto the state $\ket{\psi^\perp}=\sin(\gamma)\ket{0}-
\cos(\gamma)\ket{1}$ is obtained, giving an operation proportional to
\begin{equation}
\label{eq:fail}
\cos(\beta)\sin(\gamma)\otimes\openone^{\otimes N}
- i\sin(\beta)\cos(\gamma)\otimes\sigma_z^{\otimes N}.
\end{equation}

The measurement in Step 4 is chosen such that, for the correct measurement
result, the unitary $U(\alpha)$ is applied. This will be obtained provided
\begin{equation}
\tan(\beta)\tan(\gamma) = \tan(\alpha).
\end{equation}
For the incorrect measurement result, the unitary is of the form $U(\alpha')$,
but with $\alpha'\ne\alpha$. To obtain the correct result, we now wish to
implement the unitary \mbox{$U(\alpha-\alpha')$}. Combined with the previous
incorrect unitary $U(\alpha')$, this will result in the desired $U(\alpha)$. To
achieve this, one requires a new resource state $\ket{\psi(\beta')}$ (where
$\beta'$ is not in general equal to $\beta$) and repeats the process.

If the incorrect measurement result is obtained repeatedly, one can obtain the
correct unitary by using the resource state $(\ket{0}^{\otimes N}+
\ket{1}^{\otimes N})/\sqrt{2}$. One can deterministically perform any unitary of
the form $U(\alpha)$ with this resource state. To see this, consider steps
1 to 4 with $\beta=\pi/2$ and $\gamma=\alpha$. Then, in the case of failure one
simply applies the local unitary $U(\pi/2)$.

In the following the terminology ``stage'' is used to mean steps 1 to 4. The
notation $\beta_l$ is used for the parameter for the entangled resource state
used in stage $l$ (i.e.\ after $l-1$ failures). In addition, $\alpha_l$ is used
for the parameter for the unitary we wish to implement in stage $l$. That is,
$\alpha_1=\alpha$, $U(\alpha_2)$ is the unitary we require for the correction
after one failure, and so forth. We also use $L$ for the maximum total number of
stages.

If $N=2$, $\alpha=\pi/2n$ and $\beta_l=\alpha_l$, this method is similar to that
of CDKL. At each stage the probability of success is equal to 1/2, and for
repeated failure we need to implement $U(\pi/2)$, which may be performed
locally. The resource state is different from that used by CDKL, but the average
entanglement consumed is identical.

\begin{figure}[t]
\centering
\includegraphics[width=0.45\textwidth]{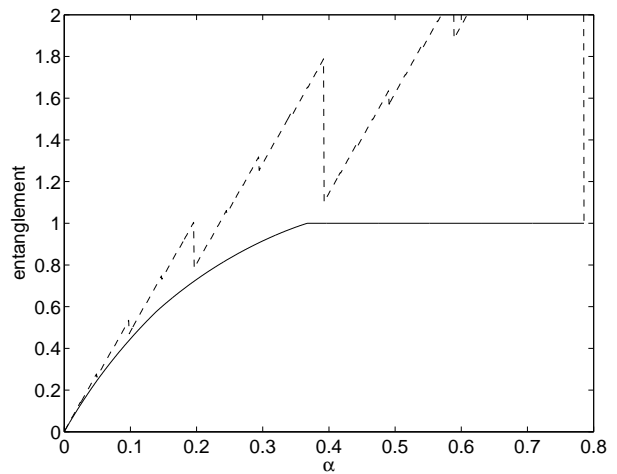}
\caption{The average entanglement consumed to implement the operation
$U(\alpha)$ as a function of $\alpha$. The case where the resource states have
been numerically optimised is shown as the solid line, and the case for the
scheme of Ref.\ \cite{cirac} is shown as the dashed line.} \label{fig0}
\end{figure}

To improve upon the entanglement consumption we adjust the entanglement of the
resource states. By using resource states with larger entanglement, the
probability for success is increased, and the average entanglement consumed is
reduced. It is straightforward to numerically optimise for the resource states
that minimise the average entanglement consumed for given $\alpha$. The
resulting average entanglement is plotted in Fig.\ \ref{fig0} as a function of
$\alpha$.

The entanglement consumed for the CDKL scheme is also shown in Fig.\ \ref{fig0}.
The CDKL scheme is explicitly given for values of $\alpha$ of the form $\alpha=
\pi/2n$. However, it is trivial to apply it to other values of $\alpha$ by
making an expansion $\alpha=\sum_n a_n \pi/2^n$, with $a_n\in\{0,1\}$. The
entanglement consumed may then be determined by summing it over that for the
individual terms. This is how the values shown in Fig.\ \ref{fig0} were
determined.

For all values of $\alpha$ (except the trivial $\alpha=0$) the entanglement
consumed for the numerically optimised scheme is less than that for the CDKL
scheme. For small $\alpha$ the entanglement consumed is approximately
$5.6418\alpha$, as compared to $5.9793\alpha$ for the CDKL scheme.

To show that the entanglement consumed is approximately $5.6418\alpha$ in the
limit of small $\alpha$, we may use the following approach. First we determine
the entanglement required to implement $U(\Alpha)$ for values of $\Alpha$ over a
range of a factor of two. Here we take $\Alpha\in [\pi/2^{20},\pi/2^{19})$. Now
for values of $\alpha$ below $\pi/2^{20}$, we select an $\Alpha$ in the range
$[\pi/2^{20},\pi/2^{19})$, such that $\alpha=\Alpha/2^n$. Then, rather than
numerically optimising for the best intermediate entangled states, we select the
intermediate entangled states with $\beta_l=\alpha 2^{l-1}$ for $1\le l \le n$,
so $\beta_l=\alpha_l$. If a successful measurement has not been obtained before
stage $n+1$, then the correction required is $U(\Alpha)$; this may be achieved
using the numerically optimised scheme.

The average entanglement consumed is
\begin{equation}
\label{eq:up1}
\E(\alpha)=\E(\Alpha)2^{-n} + \sum_{l=1}^n 2^{1-l} E(\alpha_l),
\end{equation}
where we use $\E(\alpha)$ for the entanglement consumed to implement
$U(\alpha)$, and $E(\alpha)$ for the entanglement of the resource state
$\ket{\psi(\alpha)}$. We have $E(\alpha)=h[\sin^2(\alpha)]$ where $h$ is the
binary entropy function $h(p)=-p\log_2(p)-(1-p)\log_2(1-p)$.

This entanglement measure is the entropy of the reduced density operator for
one subsystem. This is a consistent entanglement measure for Schmidt
decomposable multipartite pure states. It is possible to apply similar methods
to the bipartite case to perform entanglement concentration and dilution
between these states and the standard states $(\ket{0}^{\otimes N}+
\ket{1}^{\otimes N})/\sqrt{2}$ \cite{bennett}.

The first term in Eq.\ \eqref{eq:up1} is the probability for $n$ failures
($2^{-n}$) multiplied by the entanglement required to implement $U(\Alpha)$.
The sum is the entanglement of the state with $\beta=\alpha_j$ for steps 1 to
$n$ multiplied by the probability. Eq.\ \eqref{eq:up1} gives the upper bound
\begin{equation}
\label{eq:up2}
\frac{\E(\alpha)}{\alpha}
\le \frac{\E(\Alpha) + \sum_{k=1}^\infty 2^{k} E(\Alpha 2^{-k})}{\Alpha}.
\end{equation}

\begin{figure}[t]
\centering
\includegraphics[width=0.45\textwidth]{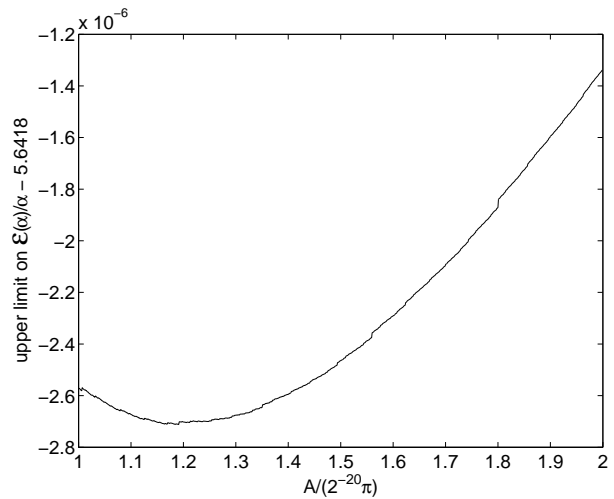}
\caption{Upper limit on $\E(\alpha)/\alpha$ [the RHS of Eq.\ \eqref{eq:up2}]
minus 5.6418 as a function of $\Alpha$.} \label{rat}
\end{figure}

The expression on the right-hand side (RHS) of Eq.\ \eqref{eq:up2} is
independent of $n$, and only depends on $\Alpha$. This expression is plotted
as a function of $\Alpha$ in Fig.\ \ref{rat} (the irregularity appears to be
due to finite precision). It can be seen that this expression does not exceed
$5.6418$ for this range of $\Alpha$. Thus we find that, in the limit of small
$\alpha$, the entanglement consumed need not exceed $5.6418\alpha$.

This proof is not fully rigorous because it is not possible to determine the RHS
of Eq.\ \eqref{eq:up2} for all values of $\Alpha$ in this range. For rigour,
note that $\alpha$ may be expanded as $\alpha=\sum_n a_n \pi/2^n$, and the
simulation of $U(\alpha)$ may be achieved by simulation of the appropriate
$U(\pi/2^n)$. In practice, this method is more complicated, however.

Another issue is that the above analysis is in terms of the expectation value of
the entanglement consumed, given that one is provided with entangled states on
demand. Alternatively we may consider many implementations of the unitary with
a fixed set of entangled resource states. This is consistent with the definition
given in Sec.~\ref{sec:gen}.

Let $p(l)$ be the probability of requiring the entangled resource state in stage
$l$ (i.e.\ the probability of $l-1$ failures). Now, for all $\epsilon,\delta>0$,
there exists an $M$ such that the probability of the number of entangled
resource states required in stage $l$ exceeding $Mp(l)(1+\delta)$ is no more
than $\epsilon/L$. Therefore, let us consider $M$ implementations of $U(\alpha)$
with $\lfloor Mp(l)(1+\delta)\rfloor$ copies of resource state
$\ket{\psi(\beta_l)}$ at each stage. Then the total entanglement does not exceed
$M\E(\alpha)(1+\delta)$.

In the case that the number of resource states exceeds that provided in one of
the stages, let the output state be $\rho_{\rm fail}$. Otherwise the correct
output state $\ket\phi$ is obtained. After discarding the measurement results,
the average output state is of the form
\begin{equation}
\rho=p_{\rm fail}\rho_{\rm fail}+(1-p_{\rm fail})\ket\phi\bra\phi.
\end{equation}
Because the total probability of requiring more than the number of resource
states provided in stage $l$ is no more than $\epsilon/L$, the total probability
of failure does not exceed $\epsilon$. Therefore the fidelity must be at least
$1-\epsilon$. Thus we find that, given sufficiently large $M$, the average
entanglement consumed is arbitrarily close to $\E(\alpha)$ with fidelity
arbitrarily close to 1.

\section{Classical communication}
\label{sec:clascom} The next issue to consider is that of the classical
communication required. At stage $l$ the set of entangled resource states is of
the form
\begin{equation}
\ket{\Psi_l}=[\cos(\beta_l)\ket{0}^{\otimes N}+i\sin(\beta_l)
\ket{1}^{\otimes N}]^{\otimes M_l},
\end{equation}
where $M_l$ is the number of entangled states used in stage $l$, and is equal to
$\lfloor Mp(l)(1+\delta)\rfloor$. We may alternatively express this state as
\begin{equation}
\ket{\Psi_l}=\sum_{\I} \mu_{\I}\ket{\I},
\end{equation}
where $\I=(i_1,\ldots,i_{M_l})$, $\mu_{\I}=\prod_m \sin^{i_m}(\beta_l)
\cos^{1-i_m}(\beta_l)$ and $\ket{\I}=\ket{i_1}\otimes\ldots\otimes
\ket{i_{M_l}}$.

Now we make an approximation to the state $\ket{\Psi_l}$ by retaining only
typical sequences of the $i$. Denoting the set of typical $\I$ by $S_l$, the
approximate state is
\begin{equation}
\ket{\tilde\Psi_l}\propto\sum_{\I\in S_l} \mu_{\I}\ket{\I}^{\otimes N}.
\end{equation}
It is a standard result for typical sequences that, for all $\epsilon,\delta>0$,
there exists an $M_l$ such that the fidelity is at least $1-\epsilon/L$ with the
number of elements in $S_l$ no more than $2^{M_lE(\beta_l)(1+\delta)}$.

Because fidelity does not decrease under completely-positive trace-preserving
(CPTP) maps, the fidelity of the final state with the output for the exact state
must be at least $1-\epsilon/L$. In the case where we use approximate resource
states with typical sequences at each stage, the final output must have fidelity
at least $1-\epsilon$ with the state obtained with the exact resource states,
$\rho$. The fidelity with the desired state, $\ket\phi$, is then at least
$1-2\epsilon$.

Now, to reduce the average communication required, we replace Step 2 with a
joint measurement in the Fourier transform basis on the states $\ket{\I}$ for
typical sequences by parties 1 to $N-1$. The communication required for the
measurement result does not exceed $M_lE(\beta_l)(1+\delta)$ bits. Using the
expression for $M_l$, this does not exceed $Mp(l)E(\beta_l)(1+\delta)^2$.

There are a number of different ways this communication may be performed. The
communication may be performed from party 1 to 2, then the sum (modulo
$\|S_l\|$) communicated to party 3, and so forth up to the communication to
party $N$. Alternatively the communication may be performed directly from each
party (1 to $N-1$) to party $N$. Many other combinations of communication are
possible, provided it is possible for party $N$ to determine the sum (modulo
$\|S_l\|$) of the measurement results.

After this communication is performed, we simply perform Step 4 as above. Now it
is necessary to communicate the positions of the measurement successes from
party $N$ to the other parties. It does not appear to be possible to reduce this
communication via the typical sequences approach as it does for the other
communication.

The average total communication from each party 1 to $N-1$ does not exceed
$\sum_l p(l)E(\beta_l)(1+\delta)^2 = \E(\alpha)(1+\delta)^2$. Thus, in the limit
of small $\delta$, this communication is the same as the entanglement consumed.
The communication from party $N$ will, in general, be much larger, and does not
scale down with $\alpha$. For example, in Fig.\ \ref{fig0} the communication
required is over 9 bits for the smallest (nonzero) value of $\alpha$ plotted.

\begin{figure}[t]
\centering
\includegraphics[width=0.45\textwidth]{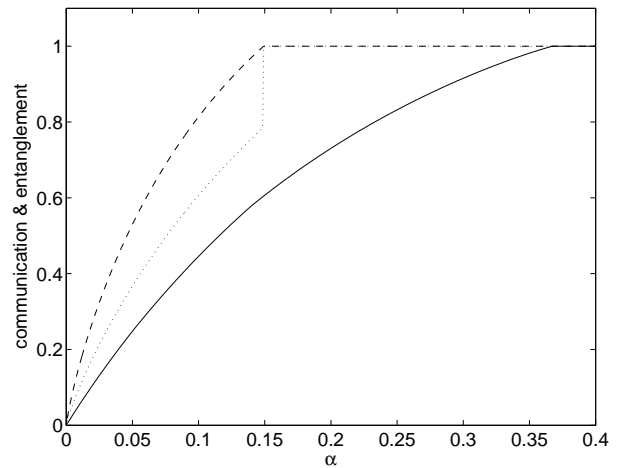}
\caption{Results for schemes optimised for minimum classical communication from
the last party. This communication is shown as the dashed line, the
corresponding entanglement required is shown as the dotted line. The
entanglement for the entanglement-optimised scheme is shown as the solid line
for comparison.} \label{comopt}
\end{figure}

It is also possible to select the entangled resources to minimise the average
communication required from the last party. In this case numerical results
indicate that the optimal scheme uses an entangled resource with
$\tan(\beta)=\tan^2(\alpha)$. For small $\alpha$, the probability of failure is
small. In the case of failure we simply implement the scheme with 1 ebit of
entanglement and 1 bit of communication in each direction. The corresponding
classical communication is shown in Fig.\ \ref{comopt}.

\begin{figure}[t]
\centering
\includegraphics[width=0.45\textwidth]{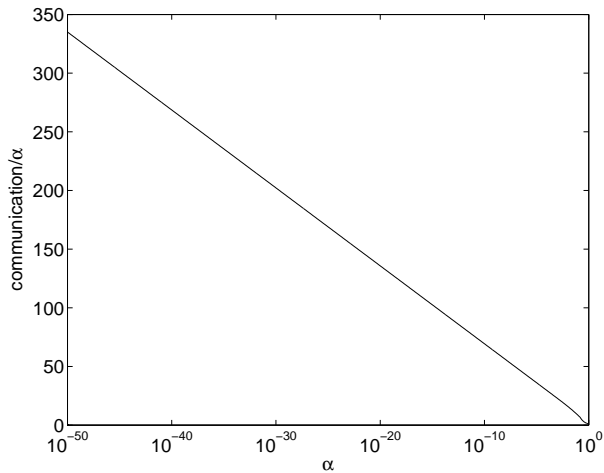}
\caption{The ratio of the communication from the final party to the interaction
strength $\alpha$.} \label{unbou}
\end{figure}

The classical communication is still larger than the entanglement for the scheme
optimised for minimum entanglement consumption. However, it does go to zero in
the limit of small $\alpha$. Unfortunately the ratio of the classical
communication to $\alpha$ is not bounded for small $\alpha$ (see Fig.\
\ref{unbou}). The ratio increases approximately as $|\log\alpha|$. This means
that a scheme based on implementing a series of operations of the form
$U(\alpha)$ will still require a large amount of communication from the last
party.

\section{General tensor product Hamiltonians}
\label{sec:genten} In the previous sections we have shown how to implement
evolution under a Hamiltonian of the form $\sigma_z^{\otimes N}$. Now we show
how to implement a general tensor product Hamiltonian of the form
$H=H_1\otimes\ldots\otimes H_N$. To do this, we apply a method similar to that
in Ref.\ \cite{child}. First note that it is possible to diagonalise the $H_j$
via local unitaries. Taking $\Delta=\|H\|$, the diagonalised form of $H$ is
\begin{equation}
H_{\rm diag}=\Delta\bigotimes_j{\rm diag}(a_{1,j},a_{2,j},a_{3,j},\ldots,
a_{d_j-1,j},a_{d_j,j}).
\end{equation}
where $a_{l,j}\in[-1,1]$, and $d_j$ is the dimension of the subsystem which
$H_j$ acts upon. In the following we consider simulation of this diagonalised
form.

This Hamiltonian may be simulated by the Hamiltonian $\sigma_z^{\otimes N_k}$
in a similar way as in Ref.\ \cite{child}. The chain of simulations used is
\begin{align}
\label{chain} &\sigma_z^{\otimes N} \to H''_{A_1B_1} \otimes \sigma_z^{\otimes
N-1} \to H''_{A_1B_1} \otimes H''_{A_2B_2} \otimes\sigma_z^{\otimes N-2} \nn &
\ldots\to \bigotimes_j H''_{A_jB_j}\to H_{\rm diag}/\Delta,
\end{align}
where $H''_{A_jB_j}={\rm diag}(a_{1,j},-a_{1,j},\ldots,a_{d_j,j},-a_{d_j,j})$.
We use the $A_j$ subscripts to indicate the subsystems upon which we wish to
implement $H$, and $B_j$ to indicate ancilla subsystems.

In order to perform this chain of simulations, we need to, in general, perform
the simulation
\begin{equation}
\bigotimes_{n=1}^{j-1} H''_{A_nB_n}\otimes\sigma_z^{\otimes (N-j+1)}
\to\bigotimes_{n=1}^{j} H''_{A_nB_n}\otimes\sigma_z^{\otimes(N-j)}.
\end{equation}
To do this, we first append the $d_j$ dimensional ancilla $B_j$ so the $j$'th
term in the tensor product is ${\rm diag}(1,-1,1,-1,\ldots,1,-1)$. Now we take
$p_l=(a_{l,j}+1)/2\in[0,1]$, and define the local unitaries $U_l$ which exchange
the $(2l-1)$-th and $(2l)$-th basis vectors of $A_jB_j$. To simulate
$\bigotimes_{n=1}^{j} H''_{A_nB_n}\otimes\sigma_z^{\otimes(N_k-j)}$ for small
time $\delta t$, we simply apply $U_l$ at time $p_l\delta t$ and again at time
$\delta t$. In this way, we may apply each successive simulation in the chain
\eqref{chain}. The final simulation in the chain may be achieved by simply
restricting to the appropriate subspace.

Ref.\ \cite{child} gave a similar method for the case of bipartite tensor
product Hamiltonians (though the multipartite case was mentioned briefly in the
discussion). In the bipartite case of Ref.\ \cite{child}, the simulation is
reversible. This is because the diagonal Hamiltonian may be expanded as a sum of
local Hamiltonians and a Hamiltonian for which the maximum and minimum
eigenvalues have the same magnitude. The local Hamiltonians may be ignored,
because they can be implemented locally. One can then use $H_{\rm diag}/\Delta$
to simulate $\sigma_z^{\otimes N}$ by restricting to the subspace for the
maximum and minimum eigenvalues. In the multipartite case we can not use a
similar method, because the additional terms will be multipartite Hamiltonians
on $N-1$ subsystems, rather than local Hamiltonians.

The case of self-inverse Hamiltonian evolution, as in Ref.\ \cite{zeng}, is
particularly simple. All the $a_{l,j}$ are $\pm 1$, so it is possible to achieve
the entire simulation chain without intermediate unitaries simply by restricting
to the appropriate subspaces. Thus self-inverse Hamiltonian evolution is
equivalent to evolution under $\sigma_z^{\otimes N}$.

In order to simulate evolution under $H$ for a time interval $t$, we divide the
time interval into $m$ intervals of length $t/m$. For $m$ sufficiently large,
the chain of simulations above will be accurate. The expectation value for the
consumed entanglement is then no more than $5.6418t\|H\|$. In fact, the
expectation value is approximately equal to this, because each simulation of
$U(\alpha)$ is for $\alpha\ll 1$. As above, we may also obtain an asymptotic
average entanglement consumed equal to this value in the limit of a large number
of implementations of $H$. Also the asymptotic average communication from each
of the parties $1\ldots N-1$ may be made equal to this value.

Note that it is necessary to consider a large number of implementations of the
Hamiltonian for small time intervals, because in addition to the division of $t$
into $m$ intervals in order to make the simulations in Eq.\ \eqref{chain}
accurate, the simulations in Eq.\ \eqref{chain} require further subdivision of
the time. However, for a given fidelity $1-\epsilon$ required, the number of
subintervals is fixed, so there is no problem in taking the appropriate limits
to obtain the average entanglement and communication.

We can also use the above methods to simulate more general Hamiltonians of the
form $H=\sum_k H_k$, where each $H_k$ is a tensor product Hamiltonian. We simply
implement $(\Pi_k e^{-itH_k/m})^m$. For large $m$, this simulation may be made
arbitrarily accurate. The average entanglement consumed is approximately
$5.6418t\sum_k\|H_k\|$. For completely general Hamiltonians a large number of
terms in the sum will be required, so this method will not be efficient.
However, it should be useful for cases with only a moderate number of terms
required.

\section{Further generalisation}
\label{sec:fur} We may generalise the above methods to general unitaries of the
form
\begin{equation}
\label{unitex}
U = \sum_k \lambda_k V_k^{(1)} \otimes V_k^{(2)} \otimes \ldots \otimes
V_k^{(N)},
\end{equation}
where the $V_k^{(j)}$ are local unitaries, and $\lambda_k$ is small except for
the $k$ such that $V_k^{(j)}=\openone$.
To implement this operation, we use an entangled resource state of the form
\begin{equation}
\ket{\Psi_U} = \sum_k \mu_k \ket{k}_{B_1} \otimes \ldots \otimes \ket{k}_{B_N}.
\end{equation}
We use $B_j$ for the subsystems for the resource state, and $A_j$ for the
subsystems for the system state (which we wish to apply $U$ to).

Each party $j$ performs the controlled operation $\sum_{k} \ket{k}_{B_j}\bra{k}
\otimes V_k^{(j)}$, where $V_k^{(j)}$ acts upon $A_j$. As a result, we have the
stator
\begin{equation}
\sum_k \mu_k \ket{k}_{B_1}\otimes V_k^{(1)} \otimes \ldots \otimes \ket{k}_{B_N}
\otimes V_k^{(N)}
\end{equation}
acting upon the system.

Each party performs local measurements, and communicates the results to party
$N$, where the phase correction is made. We then perform a projective
measurement on the resource state in subsystem $B_N$, with success corresponding
to projection onto the state $\sum_k \nu_k \ket{k}$. Choosing
$\mu_k\nu_k^*\propto\lambda_k$, $U$ is implemented in the case of success.

In the case of failure a different unitary operation is performed. This case may
be corrected for by implementing the correction in the same way, or
alternatively by teleporting the portions of the state in subsystems $A_1$ to
$A_{N-1}$ to party $N$, performing the appropriate unitary, then teleporting
these portions of the state back to their respective subsystems.

For example, we may use this approach with two-qubit unitaries. Two-qubit
unitaries may be simplified to the form \cite{khaneja,nielsen}
\begin{equation}
U=\sum_{k=0}^3 \lambda_k \sigma_k\otimes\sigma_k,
\end{equation}
where we take the convention that $\sigma_0=\openone$. In the case of the
incorrect measurement result, the error operation is again of this form, and we
wish to implement an operation of this form for the correction. Repeating the
process, we may implement the unitary in a more direct way than by implementing
a sequence of unitaries of the form $U(\alpha)$. However, calculations indicate
that this method consumes more entanglement.

One use of this approach would be to implement the operation with small
classical communication in all directions. As discussed at the end of Sec.\
\ref{sec:clascom}, it is possible to optimise for minimum communication from the
last party, however the ratio of the classical communication to $\alpha$ is not
bounded. This means that, if we use this method for more general tensor product
Hamiltonians, the average communication from the last party is still going to be
large. On the other hand, using the approach given in this section, we can
ensure that the average communication required approaches zero in the limit as
$U$ approaches the identity. By selecting $\mu_k\propto\nu_k\propto
\sqrt{\lambda_k}$, we find that the probability of failure approaches zero as
$U$ approaches the identity.

\section{Conclusion}
\label{sec:con} We have presented a scheme for implementing evolution under
general multipartite tensor product Hamiltonians. This scheme requires
multipartite entanglement of approximately $5.6418t\|H\|$. In contrast, the
scheme of CDKL \cite{cirac}, which applies to the $\sigma_x\otimes\sigma_x$
interaction, requires entanglement of $5.9793t\|H\|$. This scheme may be applied
both to the case of a single implementation, in which case this is the
expectation value of the entanglement consumption, or to multiple
implementations.

In the case of multiple implementations, the entanglement consumption is the
average over the implementations, given that the implementations are achieved
with arbitrarily high fidelity. In addition, for multiple implementations
the average communication in some of the directions may be reduced to be equal
to the entanglement consumption. However, a large amount of communication from
one of the subsystems is required.

This scheme may also be applied to more general Hamiltonians. A general
Hamiltonian may be expressed as a sum of tensor product Hamiltonians, and
therefore may be implemented using the above approach together with a Trotter
expansion. The drawback to this approach is that it will be inefficient if a
large number of terms is required. It is also possible to apply this approach to
CPTP maps which are close to the identity. By adding an ancilla, the map may be
replaced with a unitary which is close to the identity, which may be implemented
in the above way. Again this method will be inefficient if a large number of
terms are required.


\begin{thebibliography}{}
\bibitem{tele} C. H. Bennett, G. Brassard, C. Crepeau, R. Jozsa, A. Peres, and
W. Wootters, \prl {\bf 70}, 1895 (1993).
\bibitem{eisert} J. Eisert, K. Jacobs, P. Papadopoulos, and M. B. Plenio, \pra
{\bf 62}, 052317 (2000).
\bibitem{cirac} J. I. Cirac, W. D\"ur, B. Kraus, and M. Lewenstein,
\prl {\bf 86}, 544 (2001).
\bibitem{gross} B. Groisman and B. Reznik, \pra {\bf 71}, 032322 (2005).
\bibitem{chen} L. Chen and Y.-X. Chen, \pra {\bf 71}, 054302 (2005).
\bibitem{zeng} H.-S. Zeng, Y.-G. Shan, J.-J. Nie, and L.-M. Kuang,
quant-ph/0508054 (2005).
\bibitem{stator} B. Reznik, Y. Aharonov, and B. Groisman, \pra \textbf{65},
032312 (2002).
\bibitem{bennett} C. H. Bennett, S. Popescu, D. Rohrlich, J. A. Smolin, and
A. V. Thapliyal, \pra {\bf 63}, 012307 (2001).
\bibitem{child} A. M. Childs, D. W. Leung, and G. Vidal,
IEEE Trans. Inf. Theory {\bf 50}, 1189 (2004).
\bibitem{khaneja} N. Khaneja, R. Brockett, and S. J. Glaser, \pra {\bf 63},
032308 (2001).
\bibitem{nielsen} M. A. Nielsen, C. M. Dawson, J. L. Dodd, A. Gilchrist,
D. Mortimer, T. J. Osborne, M. J. Bremner, A. W. Harrow, and Andrew Hines,
\pra {\bf 67}, 052301 (2003).
\end{thebibliography}
\end{document}